\documentclass{appolb}

    \usepackage{amsfonts}

\usepackage{graphicx}

\newcommand{\ba}{\begin{eqnarray}}
\newcommand{\ea}{\end{eqnarray}}
\newcommand{\be}{\begin{equation}}
\newcommand{\ee}{\end{equation}}

\newcommand{\beqs}{\begin{eqnarray}}
\newcommand{\eeqs}{\end{eqnarray}}

\begin{document}
\title{Odderon, HEGS model and LHC data %
\thanks{Presented at the International Workshop on Diffraction and low-x - 18}%
}
\author{O.V. Selyugin$^{(a)}$ and J.R. Cudell$^{(b)}$
\address{ \sl (a) BLTPh, JINR, Dubna, Russia\\
(b) STAR Institute, Li\`ege University, Belgium  }}

\maketitle
\begin{abstract}
    We show that the impact of the maximal odderon amplitude at $t=0$
    and  $\sqrt{s} = 13 $ TeV is small. We obtain a value of $\rho(t=0)$ at $\sqrt{s}= 13 $ TeV  of the order of  $0.12$.
    The real part of the odderon amplitude grows like $\log(s/s_0)$
    at high energies, and is calculated from the analytic properties of
    the amplitude. In the framework of the HEGS model, taking the same intercept for the odderon and the pomeron
    leads to a good fit of the new LHC data at  $ \sqrt{s}= 13 $ TeV.
We also show that the main effect of the odderon  can be seen in the  region of the diffraction minimum of the differential elastic cross section.  The form and energy dependence of the odderon amplitude determined in the HEGS model reproduce
     the characteristics of the diffraction minimum at $\sqrt{s}= 7, 8$ and $ 13 $ TeV.
\end{abstract}

 The fundamental measures of hadron interactions $-$ the total cross section $\sigma_{tot}(s)$ and
    the ratio of the real part to the imaginary part of the elastic hadron scattering amplitude  $\rho(s,t)$ $-$
    are obtained from the analyses of
  the differential cross section of elastic scattering \cite{Sel-t-rho,Cud-Sel-rho18}.
  The simplest phenomenological models parametrise the $t$ dependence of the elastic
scattering amplitude ${\cal A}(s,t)$ as a falling exponential .
  However, taking into account more realistic hadronic form factors for the interaction leads to a non-exponential
  behaviour in $t$. Furthermore, there can be  other physical effects
  that can change the momentum-transfer dependence of the hadron cross sections \cite{Sel-NP1}.
 \newcommand{\slhc}{s_{\small LHC}}

The new measurements of $\sigma_{tot}$ and $\rho$ performed by the TOTEM Collaboration \cite{TOTEM-17rho} at $\sqrt{s_{\small LHC}}=13\ \rm TeV$
and $t=0$
  open the possibility to observe the ``maximal odderon" (MO), i.e. a crossing-odd amplitude  with an
 asymptotic energy dependence  such that
  $Re F_{Odd}(s,t=0)  \sim  \log(s/s_{0})^2$ and $Im F_{Odd}(s,t=0) \sim  \log(s/s_{0})$,
 with $F(s,t)={\cal A}(s,t)/s$.
Martynov and Nicolescu  \cite{MN-Odd} have proposed that the difference
  between the central prediction  of the COMPETE Collaboration  $\rho(\slhc,t=0) =0.14$
  and the recent measurement  $0.10 \ {\rm -} \ 0.09$  
results from the contribution of the MO.

  To analyse the possibility of an odderon contribution to the differential cross sections at small
 $|t|$ we use the HEGS model \cite{HEGS0,HEGS1} which quantitatively  describes, with only a few parameters, the
  differential cross section of $pp$ and $p\bar{p}$
  from $\sqrt{s} =9 $ GeV up to $13$ TeV, and includes the Coulomb-hadron interference region and the high-$|t|$ region
  up to $|t|=15$ GeV$^2$. However to avoid  possible problems
 connected with the low-energy region, we consider here only the data above
  $\sqrt{s} =100$ GeV.

   The total elastic amplitude in general receives five helicity  contributions, but at
   high energy it is enough to write it as $F(s,t) =
  F^{h}(s,t)+F^{\rm em}(s,t) e^{\varphi(s,t)} $\,, where
 $F^{h}(s,t) $ comes from the strong interactions,
 $F^{\rm em}(s,t) $ from the electromagnetic interactions and
 $\varphi(s,t) $
 is the interference phase factor between the electromagnetic and strong
 interactions \cite{bethe,selmp1}.
    The Born term of the elastic hadron amplitude at large energy can be written as
    the sum of two pomeron and two odderon contributions
 \begin{eqnarray}
 F_{\mathbb{P} }(s,t) & =& \hat{s}^{\epsilon_0}\left(C_{\mathbb{P}} F_1^2(t)  \ \hat s^{\alpha' \ t} + C'_{\mathbb{P}} A^2(t) \ \hat s^{\alpha' t\over 4} \right) \\
 F_{\mathbb{O} }(s,t) & =&  i \hat{s}^{\epsilon_0+{\alpha' t\over 4}} \left( C_{\mathbb{O} }
   + C'_{\mathbb{O}}|t|/(1-r_{0}^{2} t ) \right) A^2(t).
 \end{eqnarray}
 All terms are supposed to have the same intercept  $\alpha_0=1+\epsilon_0 = 1.11$, and the pomeron
 slope is fixed at $\alpha'= 0.24$ GeV$^{-2}$.
  The model takes into account the two hadron form factors $F_1(t)$ and $A(t)$ which correspond to  the charge and matter
  distributions \cite{GPD-PRD14}. Both form factors are calculated  as first and second moments of  the same Generalised Parton Distributions (GPDs).
It has  four free parameters (the constants $C$) at high energy:
two for the two pomeron amplitudes  and two for the odderon.
The real part of the hadronic elastic scattering amplitude is determined
   through the complexification $\hat{s}=-i s$ to satisfy the dispersion relations.
   We find that the extra factor $1/(1-r_{0}^{2} t)$ is needed in the odderon case to reproduce the data,
and we also add the constant $C_{\mathbb{O}}$
to allow for a  possible odderon  contribution at $t=0$. The final elastic  hadron scattering amplitude is obtained
after unitarisation of the  Born term
through the standard one-channel eikonal representation.

\label{sec:figures}
\begin{figure}
\begin{center}
\vglue -1cm
\includegraphics[width=0.49\textwidth] {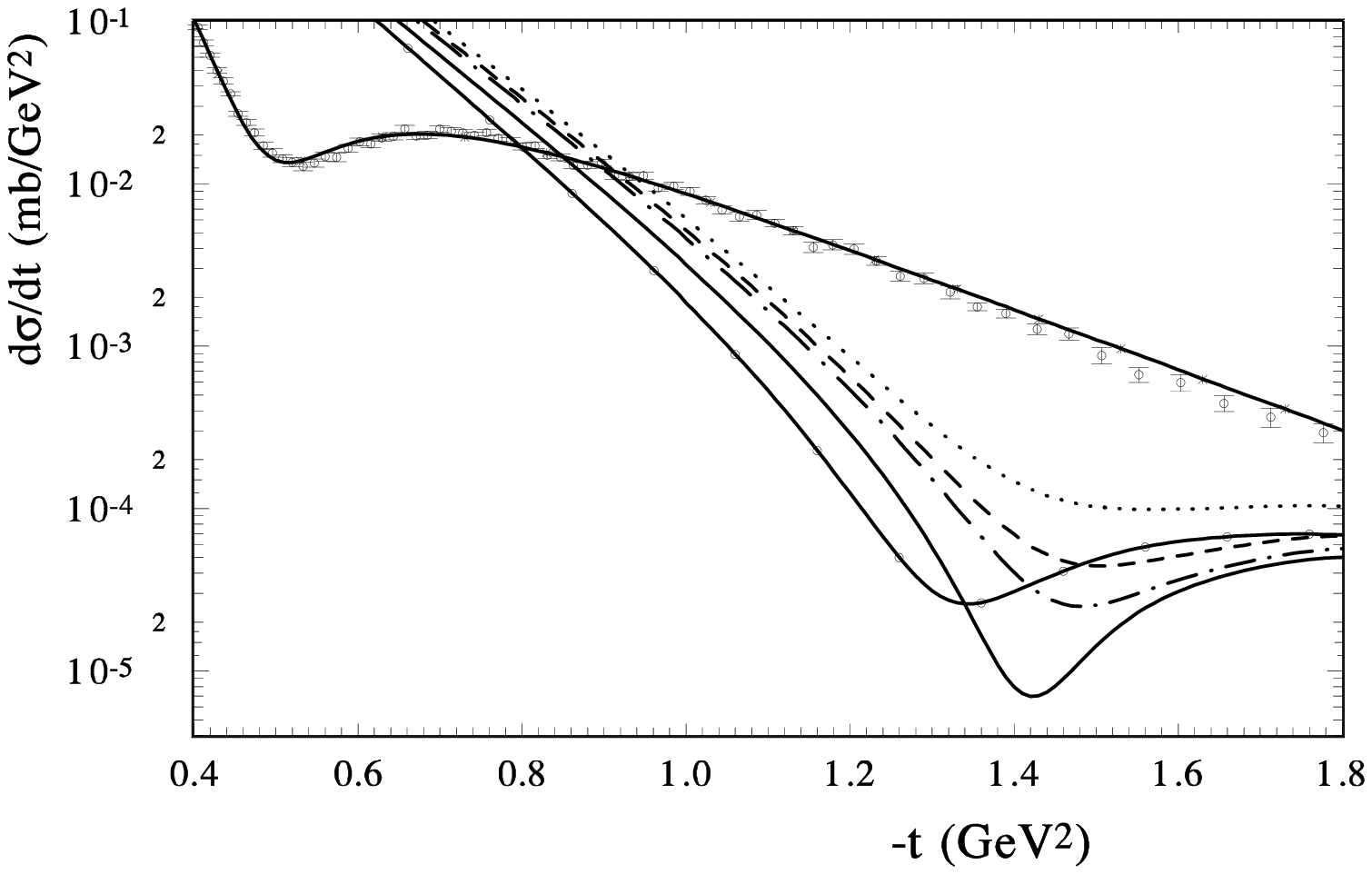}
\includegraphics[width=0.49\textwidth] {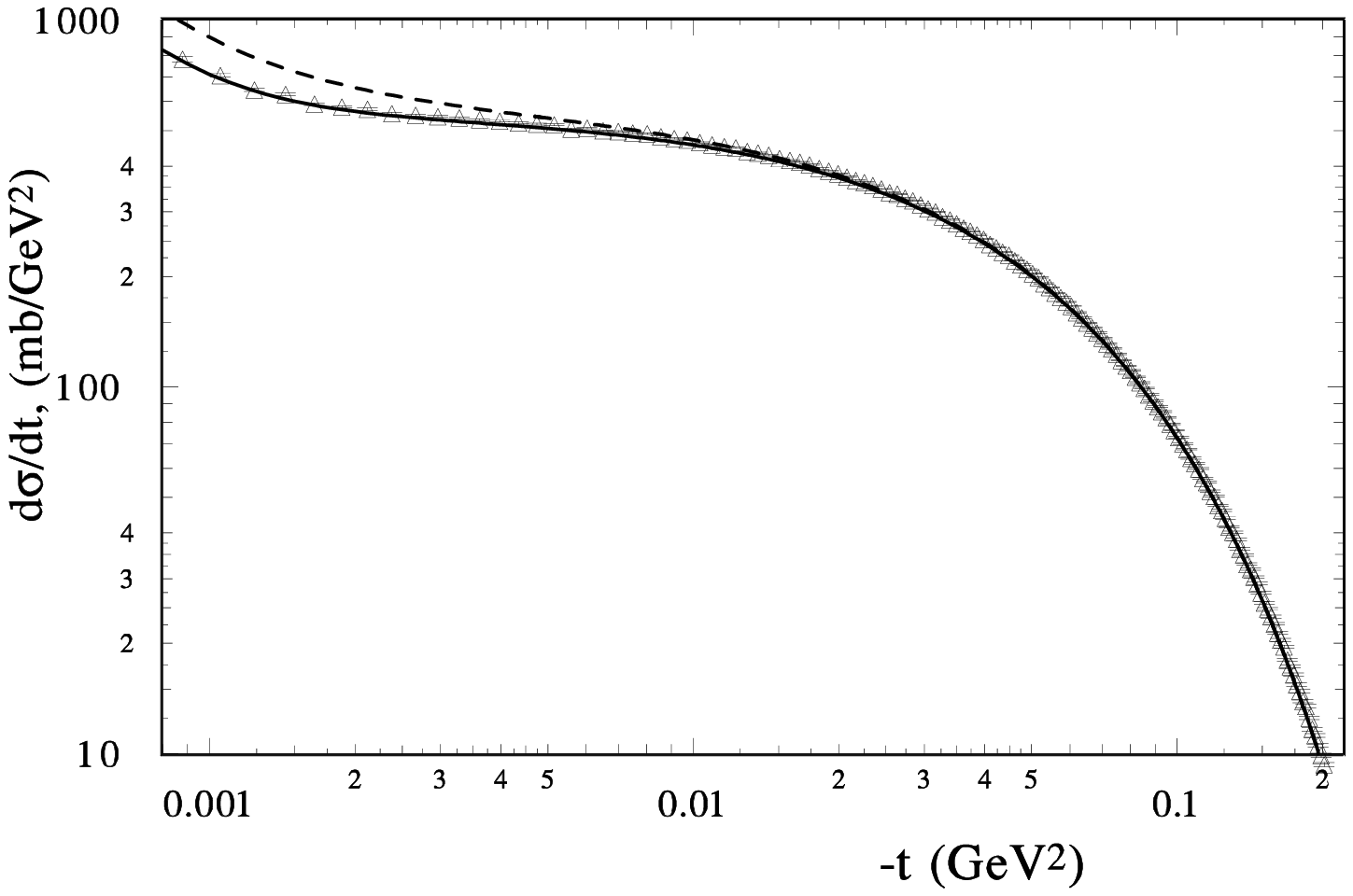}
\end{center}
\vglue 0.5cm
\caption{     $d\sigma/dt$
 [left]
  in the region of the diffraction minimum,
  lines show the HEGS results   and the experimental points are the data of the TOTEM Collaboration at $7$ TeV,
(the lines on the left correspond to  
$  \sqrt{s}=13.4; 16.8; 19.4; 30.4; 52.8; 7000  $~GeV;
  correspondingly - dots; short dash; dot-dash; solid; solid+circles; solid+ants);
  and  [right]  at small $|t|$ at $  \sqrt{s}=13.$ TeV
  (short dashed lines are for  $pp$ and dotted lines for $\bar p p$, triangles - the data of the TOTEM Collaboration at $13$ TeV )
.}
\label{Fig:d2}
\end{figure}

We include a total of $882$ experimental points for the energy region $\sqrt{s}  > 100$ GeV. In the fit, and we take into account
 only the statistical errors. The systematic errors are accounted for through
additional normalisations, one for each separate set of data.
The description of the new data on the differential cross section from  TOTEM at $\sqrt{s}=7$  and $13$ TeV
   is shown in Fig.1. The low-$|t|$   are presented with an additional normalization coefficient
   $f=0.9$.

   If we allow for the constant $C_{\mathbb{O}}$ in the odderon amplitude, we obtain  $ \chi^2 = 1132$ instead of $\chi^2= 1143$ for $C_{\mathbb{O}}=0$, and the best value is  $C_{\mathbb{O}}= -0.07 \pm 0.03$ .
    The second odderon constant is then $C'_{\mathbb{O}} = -0.29 \pm 0.01$.
Hence the odderon contributions at $t=0$ is very small and cannot heavily impact the value of $\rho(t=0)$.
Finally,
 if we neglect the odderon contribution altogether, we have only $2$ fitting parameters
 in the framework of the HEGS model. The value of $\chi^2$ slightly increases to $\chi^2= 1207$,
and the resulting fit  describes the new data of the TOTEM Collaboration
at $\sqrt{s}=13$ TeV.
\label{sec:figures}
\begin{figure}
\begin{center}
\vglue -1cm
\includegraphics[width=0.49\textwidth] {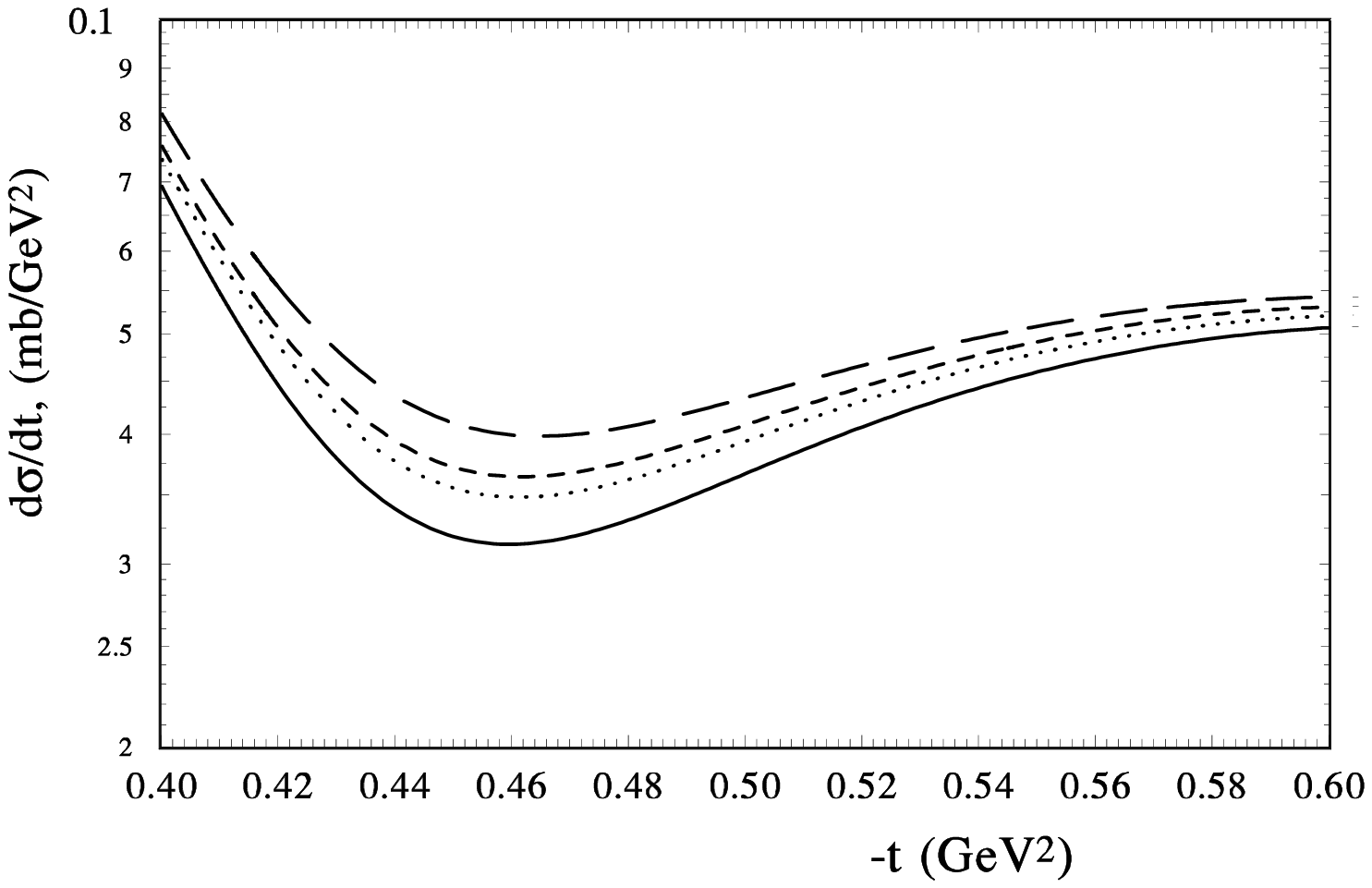}
\includegraphics[width=0.49\textwidth] {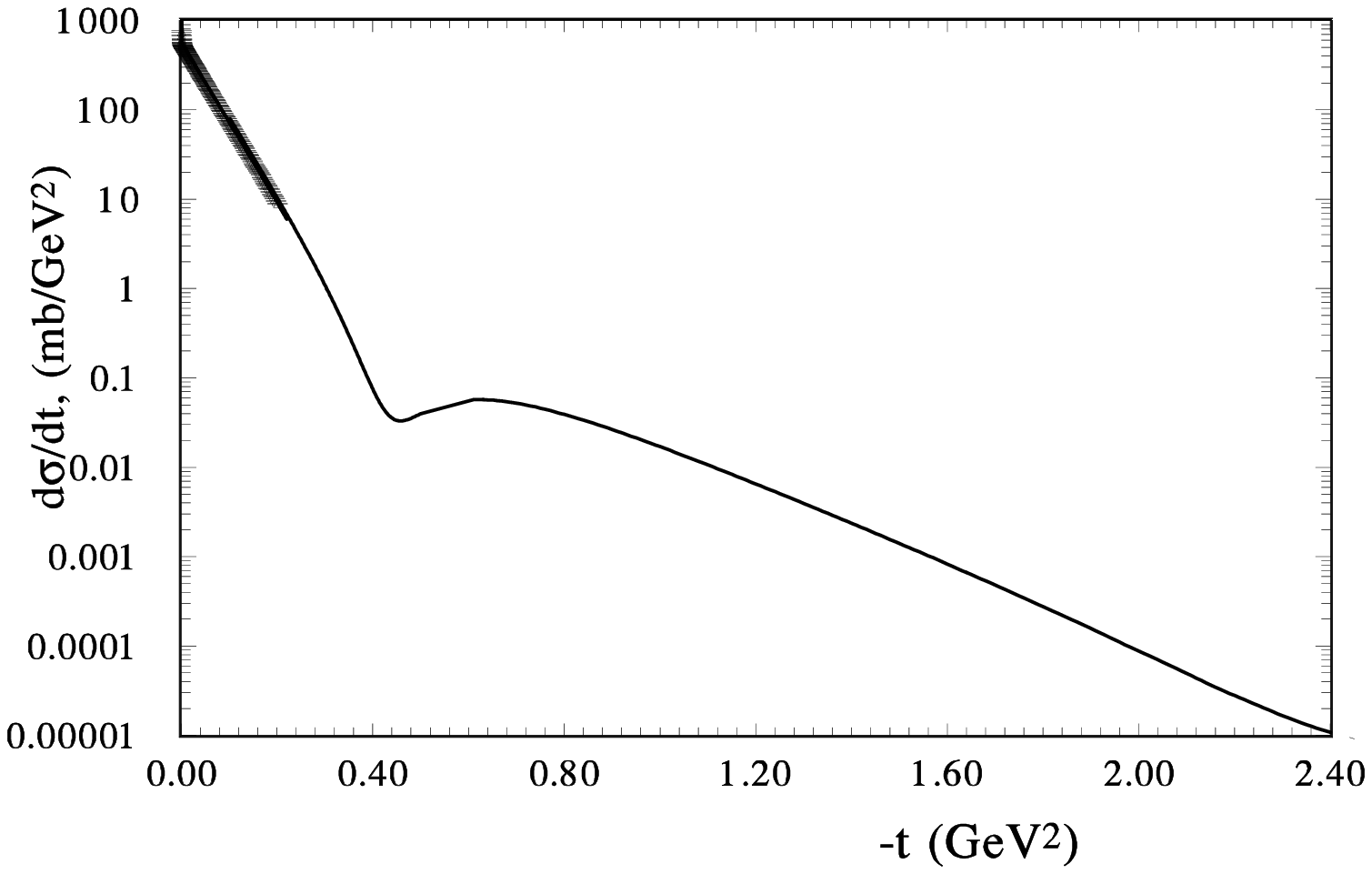}
\end{center}
\caption{
$d\sigma/dt$  for $pp$-scattering at 13 TeV,
in the region of the diffraction minimum [left] and for a wide region of $|t|$ [right]. The experimental points are the data of TOTEM.
On the left, we show the result of the HEGS calculation with odderon (plain lines and
and long-dashed lines resp. for $pp$ and $p\bar{p}$) and without odderon (short dashed line and dotted line resp. for $pp$ and $p\bar{p}$)
}
\label{Fig:d2}
\end{figure}

   In Fig. 2, the results of the HEGS model are presented at 13 TeV, near the diffraction minimum and for a wide region of
   momentum transfer.
   It is already known that the model describes the diffraction minimum and its energy dependence
  at lower energies \cite{HEGS-min}.
  As seen from the left-hand figure, for $\sqrt{s}=13$ TeV  HEGS predicts a diffraction minimum at $-t_{min}=0.46 $~GeV$^2$ and
  a maximum as $-t_{max}=0.62 $ GeV$^2$.
The maximum differential cross section is $R=1.58$ larger than the minimum one.
   This seems to agree with the latest LHC data \cite{Nemez-May18}:
   $-t_{min}=0.47 $ GeV$^2$, $-t_{max}=0.638 $ GeV$^2$ and $R=1.78$.

   The same figure also shows the differential cross section for   $\bar{p}p$. The difference between $pp$ and $\bar{p}p$ scattering
  comes from the odderon contribution. If we neglect the odderon amplitude
    the difference between $pp$ and $p\bar{p}$  scattering
is entirely  determined by the contribution from the Coulomb hadron interference and is quite small.

We can now turn to the value of $\rho$ within the  HEGS model.
    Fig. 3 [left]  shows $\rho(t)$ at $\sqrt{s}=13$ TeV for   $pp$ and $p\bar{p}$
     scattering.  At $t=0$ the best fit gives $\rho_{pp}=0.12$, i.e. only slightly less than the COMPETE central value
     \cite{COMPETE}. For $\bar{p}p$
     scattering we obtain $\rho_{\bar{p}p}=0.13$.  The difference is very small, about a fourth of that
 obtained in  \cite{MN-Odd}.
     Near $-t=0.1$, the difference $\rho_{pp}(t)-\rho_{p\bar{p}}(t)$ changes sign,
     as required by the dispersion relations \cite{Cud-Sel-DR,Cud-Sel-str}.
     At  larger $|t|$ this difference is positive.

  In Fig.3 [right], the energy dependence of  $\rho(s,t=0)_{pp}$  and $\rho(s,t=0)_{p\bar{p}}$  is shown.
   This exponential behaviour is due to the fact that we  neglected the non-asymptotic terms in the scattering
   amplitudes.
   The difference between  $\rho(t)_{pp}$ and $\rho(t)_{p\bar{p}} $ tends to zero at asymptotic $s$, if
   we set $C_{\mathbb{O}}=0$. The small value of  $C_{\mathbb{O}}$
   leads to a small additional and a small difference at asymtotic energies.
\label{sec:figures}
\begin{figure}[htb]
\begin{center}
\includegraphics[width=0.49\textwidth] {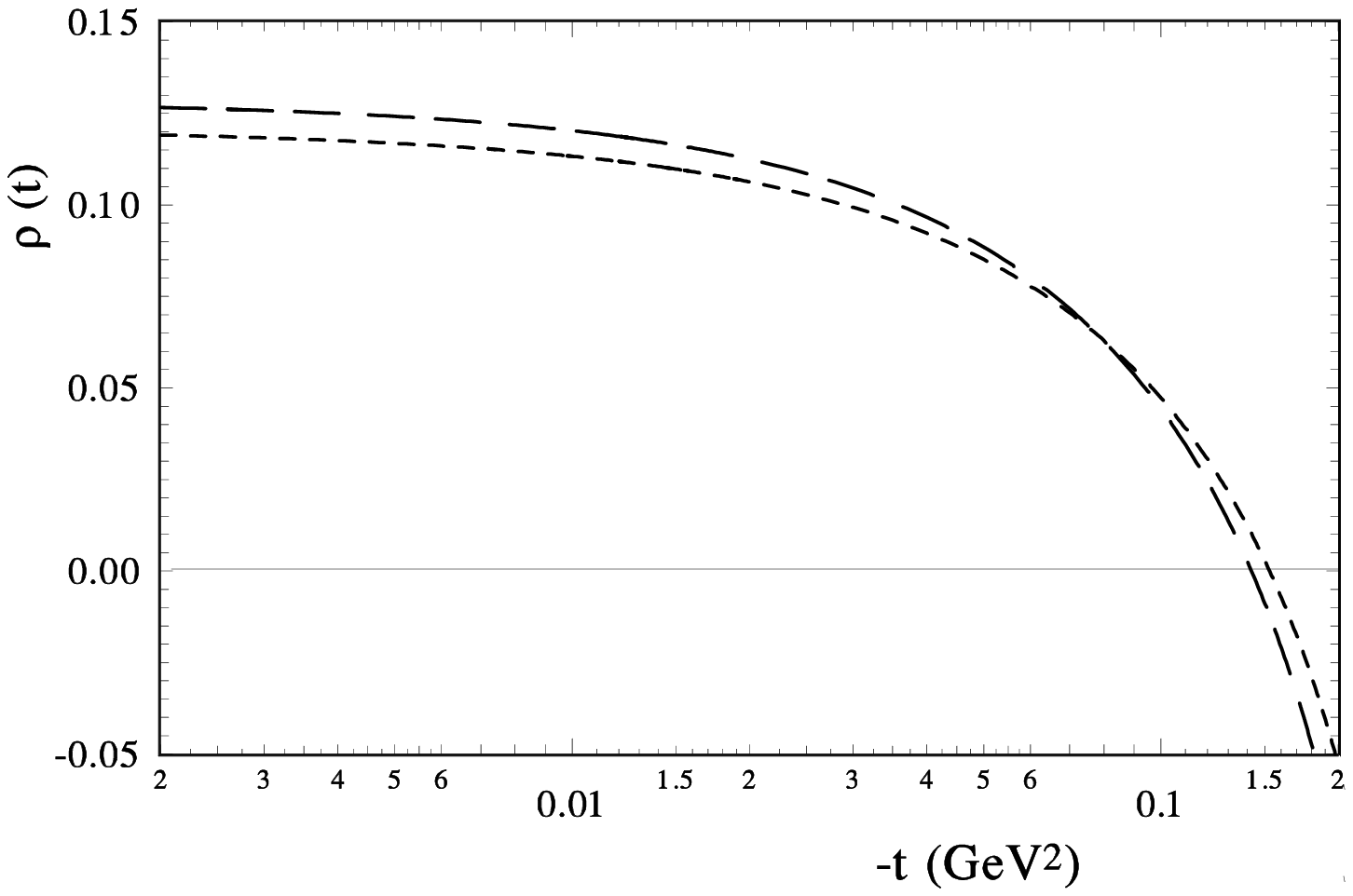}
\includegraphics[width=0.49\textwidth] {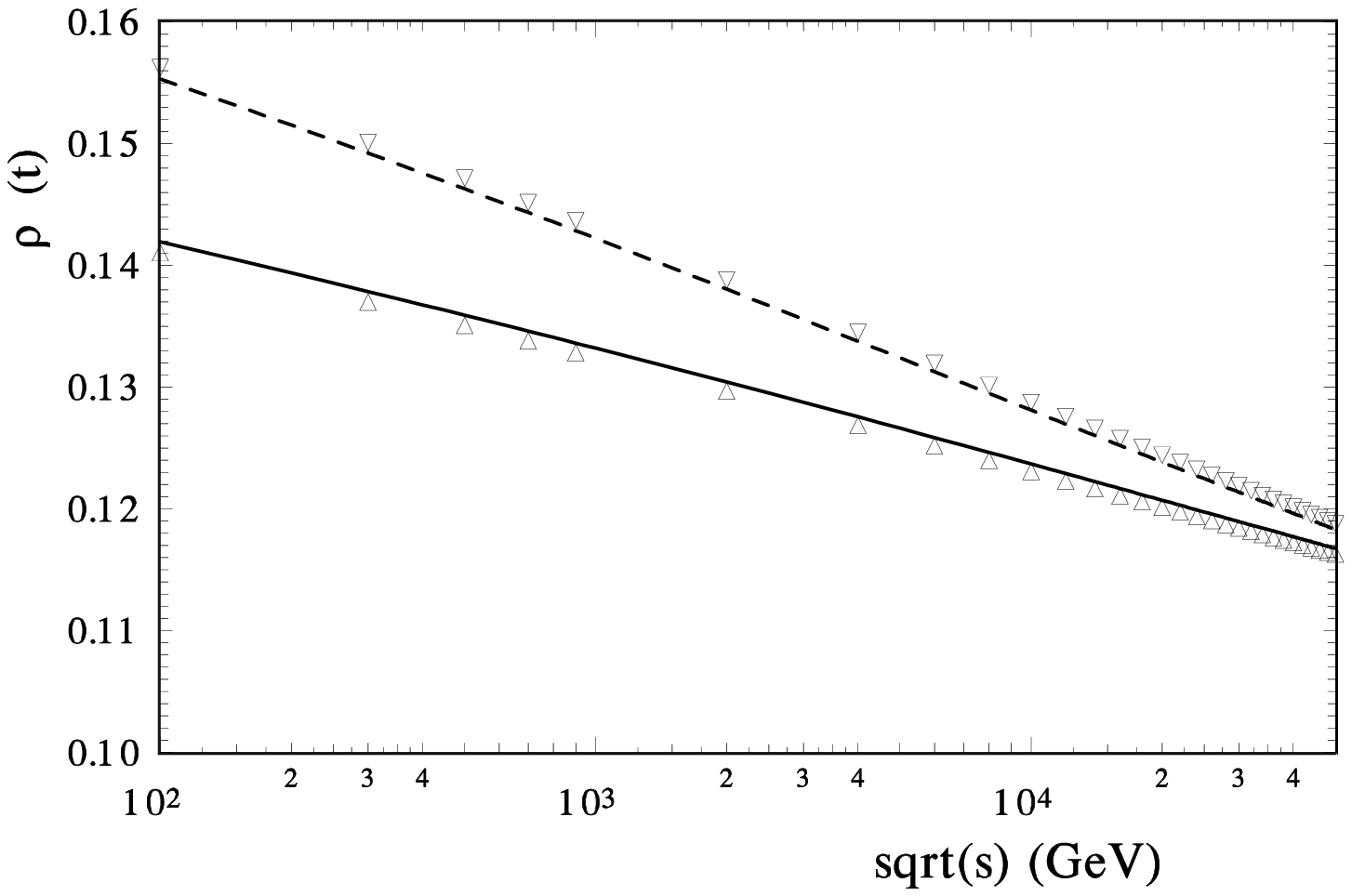}
\end{center}
\caption{  [left] $\rho(t)$ at $\sqrt{s}=13$ TeV for   $pp$ and $p\bar{p}$ scattering;
[right] The energy dependence of $\rho(s,t=0)$ for $pp$  and $p\bar{p}$ (hard and dashed lines)
    [triangle up and down show the contributions from the constant odderon term.
}
\label{Fig:d2}
\end{figure}
  The energy dependence of the scattering amplitude in the HEGS model is the same for values of $t$
  on the first diffraction cone:
  the real part grows slightly faster than $\log(s)$ but more slowly
     than $s\log^2(s)$, and the imaginary part grows slightly slower than  $s\log^2(s)$.

   The corresponding overlap function in the impact parameters representation,
   calculated in the HEGS model
 does not reach the Black disc limit (BDL) at the LHC.
The same result was obtained in the full HEGS model~\cite{HEGS1}.
   Hence the hadron interactions are not in their asymptotic regime at 13~TeV.

     Note   that  a recent paper \cite{AMT-BDL}  announced
    that  the BDL  is exceeded from the data of the TOTEM
    collaboration at $\sqrt{s}=13$ TeV. In this paper, the precision of the calculated
    $\sigma_{tot}$ and $\rho(s,t=0)$ is smaller than the precision of the experimental data by one order of magnitude.
   Their eq.(12) shows that they used the modulus of the imaginary part
     of the scattering amplitude to calculate the profile function. However the existence of a sharp
     diffraction minimum in the differential cross section at $-t=0.45$ GeV$^2$ means that the imaginary part
     changes sign in this domain of momentum transfer. The real part, which according the dispersion relations
     changes its sign in the region $-t \approx 0.1$, is larger than the imaginary part in the domain
     of the diffraction dip.  Neglecting this phenomenon may be the reason for an overestimate of the profile function.

\section*{Conclusion}
 The  analysis  of the new TOTEM data   at small momentum transfer at  $\sqrt{s}= 13 $ TeV,
    together with other experimental data at lower energies, allows to examine the energy dependence and the form
    of the odderon part of the hadron scattering amplitude.
     It has been  shown that the impact of the maximal odderon amplitude at $t=0$
    and  $\sqrt{s} = 13 $ GeV is small and cannot lead to $\rho(\sqrt{s}=13 \ {\rm GeV},t=0) = 0.09$.
The obtained value of $\rho(t=0)$ at $\sqrt{s}= 13 $ TeV is approximately equal $0.12$.
    The energy dependence of the odderon in the framework of the HEGS model with the same intercept as the pomeron
    amplitude seems to agree with the new LHC data at  $ \sqrt{s}= 13 $~TeV, if one allows for an additional normalisation
    coefficient which reflects the systematic errors.

    The form and energy dependence of the odderon amplitude determined in the HEGS model are  also
in good agreement with the features of the diffraction minimum  at $\sqrt{s}= 7,$ 8 and $ 13 $ TeV.

Hence, it is possible that the Born terms of the pomeron and odderon amplitudes have the same intercept.  The real parts of the final
scattering amplitudes both grow like $\log(s)$, as required by the analytical properties of the amplitude \cite{Fingel}.
A very different approach, using the Good-Walker formalism, leads to very similar conclusions  \cite{Khoze-1806}.

\section*{ Acknowledgements}
{\small \hspace{0.3cm} OVS would  like  to thank
Alessandro Papa and
 the Organization Committee  for the Invitation and the financial support
 and would like to thank the  University of Li\`{e}ge
  where part of this work was done. This work was also  supported
by the Fonds de la Recherche Scientifique-FNRS, Belgium, under grant No. 4.4501.15
   }

\end{document}